# Current Distribution for Superconducting Strip Transmission Lines at Microwave Frequencies.

V. M. Genkin

**Abstract**

Current distribution for a thin superconducting strip shielded by two ideally conducting plains has been calculated. It is shown that at microwave frequencies the current density has maximum over the center of the strip in contrast to the dc current pattern, which exhibits crowding over the edges.

## Introduction.

The problem of dc current distribution for a superconducting thin film has been discussed by several authors[1-5]. It was shown that dc current density in superconducting thin strip peaked at the edges. It is anticipated that the same effect takes place at microwave frequencies since the Meissner effect apply equally to ac as to dc currents. Accurate calculations of current distribution at microwave frequencies for superconducting transmission lines are very important for interpretation various experimental data and there are a lot of papers on this subject[6-12]. At microwave frequencies the distribution of the current is determined by complex interaction of the carrying current film with other conductors and crowding depends on real geometry. For real geometries of microwave devices analytic solutions for current distribution do not exist but various numerical methods have been used. In this paper we consider the current distribution for a thin superconducting strip shielded by two ideally conducting plain, i.e. stripline configuration. We found that at microwave frequency the current distribution in the superconducting strip was significantly different from dc current pattern. Current density exhibits maximum over the center of the strip.

## Problem formulation and results.

Let us consider the stripline structure of Fig. 1 with thin strip and ideal ground plains. Wave propagates along z-axis.

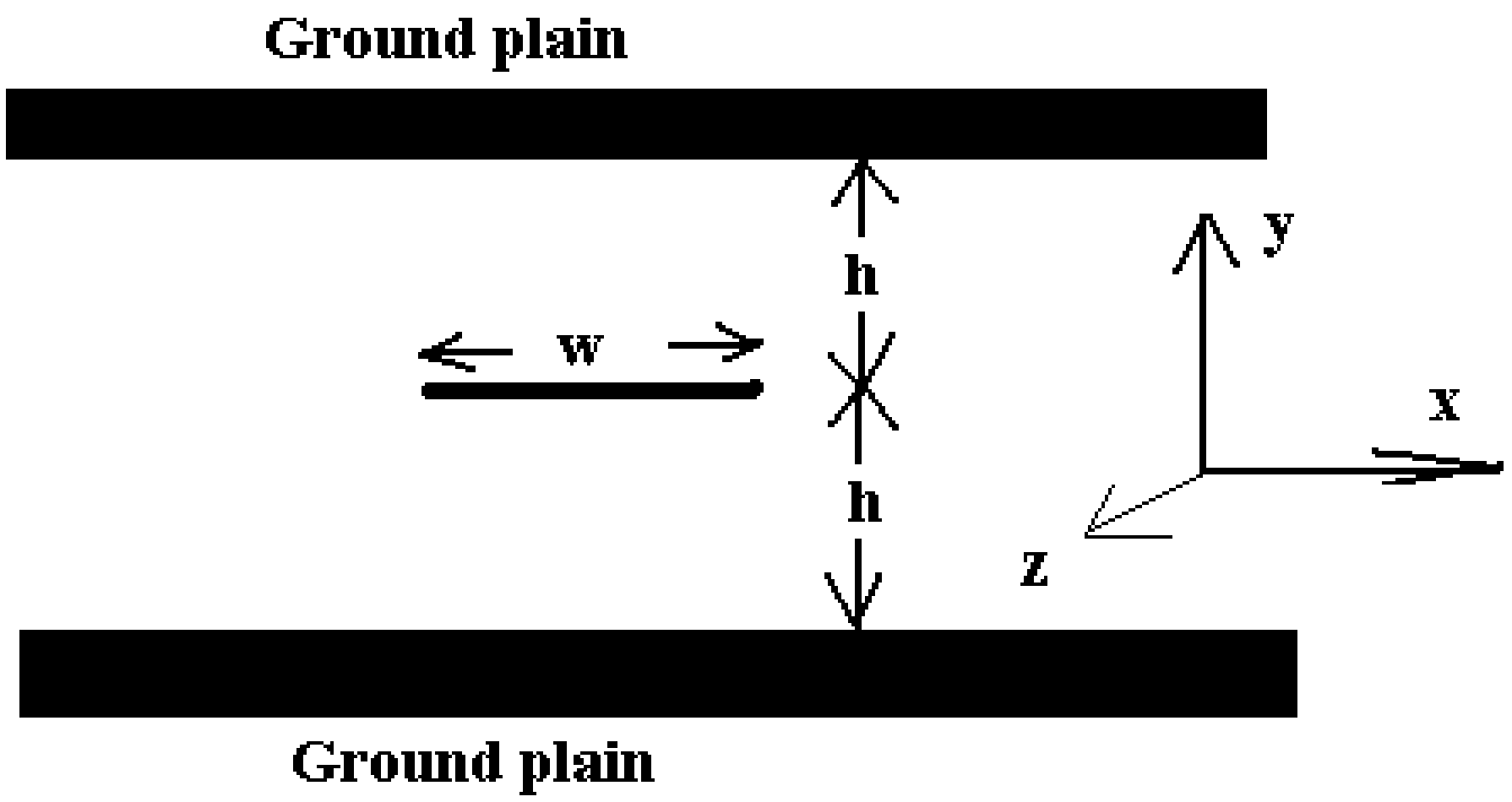

Fig. 1. A stripline structure with a thin superconducting strip.

Maxwell equations for electromagnetic fields in the form

$$\vec{E}(y,k,q)\ \exp(ikz+iqx-i\omega t),\ \vec{H}(y,k,q)\ \exp(ikz+iqx-i\omega t). \qquad (1)$$

after substitution $E_y = qH_z/k_0\varepsilon - kH_x/k_0\varepsilon$ and $H_y = kE_x/k_0 - qE_z/k_0$ are

$$k_0\varepsilon\frac{\partial E_z}{\partial y} = ikqH_z - ik_e^2 H_x$$

$$k_0\varepsilon\frac{\partial E_x}{\partial y} = ik_h^2 H_z - ikqH_x$$

$$k_0\frac{\partial H_z}{\partial y} = -ikqE_z + ik_e^2 E_x + 4\pi k_0 j_x/c \qquad (2)$$

$$k_0\frac{\partial H_x}{\partial y} = -ik_h^2 E_z + ikqE_x - 4\pi k_0 j_z/c$$

where $k_0 = \omega/c$, $k_e^2 = k^2 - k_0^2\varepsilon$, $k_h^2 = q^2 - k_0^2\varepsilon$, ε is the relative dielectric constant. Assuming that the thickness of the strip, $d$, is small in compared with the penetration length, and current density does not depend on $y$, we have such boundary conditions at $y = h$

$$E_x(h-0,k,q) = E_x(h+0,k,q),\ E_z(h-0,k,q) = E_z(h+0,k,q)$$

$$H_x(h-0,k,q) - H_x(h+0,k,q) = -4\pi d j_z(k,q)/c,$$

$$H_z(h+0,k,q) - H_z(h-0,k,q) = 4\pi d j_x(k,q)/c. \qquad (3)$$

Current density in the strip has only *x, z* components. Matching the field components we obtain for z-component of electric field at the strip

$$E_z(h,k,q) = -2\pi i d\ \tanh(\rho h)\{k_e^2 j_z(k,q) + kq j_x(k,q)\}/k_0\varepsilon\rho \qquad (4)$$

where $\rho^2 = k^2 + q^2 - k_0^2\varepsilon$. Integral equation for $E_z$ is given by

$$\int_0^w [E_{zz}(x,x')(j_z(x')+j_z^{ext})+E_{zx}(x,x')j_x(x')]dx' = j_z(x)/s \qquad (5)$$

where

$$E_{zz}(x,x') = -id\int dqk_e^2 \tanh(rh)\exp[iq(x-x')]/rk_0e$$

$$E_{zx}(x,x') = -id\int dqkq \tanh(rh)\exp[iq(x-x')]/rk_0e \qquad (6),$$

$s$ is the complex conductivity of the strip, $w$ is the strip width.. External current, $j_z^{ext}$, has been inserted to obtain integral equation of the second type. For superconducting strip this current could be ascribed to the gradient of the phase of the order parameter in accordingly with London equation

$$\vec{j} = -\frac{c}{4pl^2}\left(\vec{A} - \frac{\Phi_0 \nabla j}{2p}\right) \qquad (7)$$

where $\lambda$ is the London penetration length, $\Phi_0$ is the magnetic flux quantum. The conductivity of superconductor neglecting normal currents is $s = ic^2/4pl^2w$. The solutions of Eq. (5) with $j_z^{ext}$ =0 determine the wave numbers $k$ of propagating waves. In general, $j_x << j_z$, and we neglect *x*-component in the following qualitative discussion. The character of the solution of Eq. (5) is determined by the function $E_{zz}(x-x')$. If this function has maximum at small distances and its sign corresponds to the repulsion as for dc currents, we obtain current crowding at the edges. But if this function has minimum at small distances, then current could crowd over the center of the strip. Function $E_{zz}(x-x')$ is proportional to $k_e^2 = k^2 - k_0^2e$ and changes its sign as $k/k_0$ increases as it is demonstrated by Fig. 2 where the normalized imaginary part of $E_{zz}(x-x')$ (real part is small) is shown for $k/k_0$ =4 and $k/k_0$ =5. These calculations are performed for stripline with $h = 0.05/k_0$, and $LaAlO_3$ as a substrate ($\varepsilon$=24).

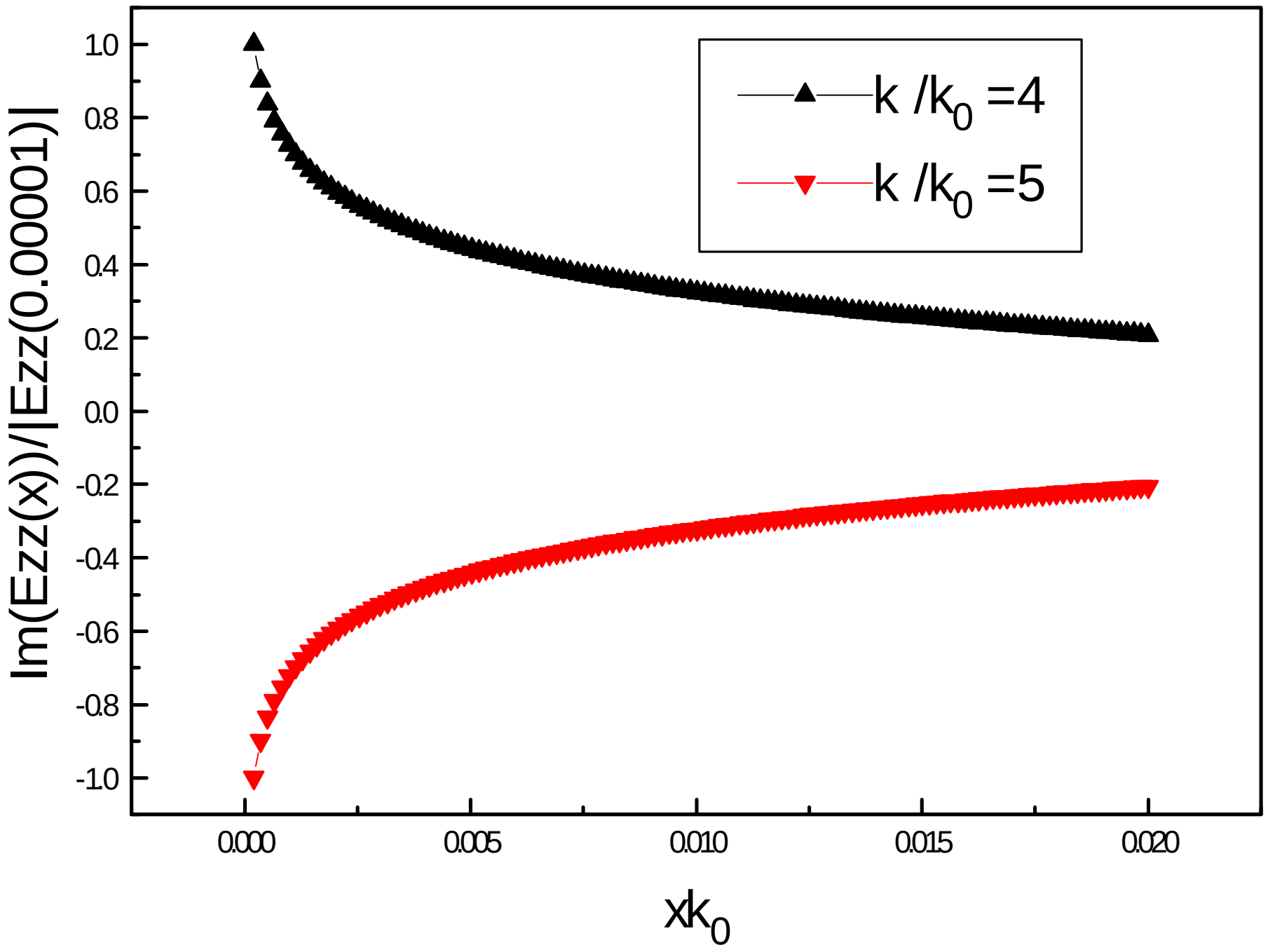

Fig. 2. Normalized imaginary part of $E_{zz}(x)$ for two values of $k/k_0$.

This changing of the function $E_{zz}(x)$ reflects in the character of the solution of Eq.(5). For $k/k_0$ =4 we obtain maximum of the current density at the strip edges; for $k/k_0$ =5 this maximum moves to the strip center. This is demonstrated by Fig. 3. There we show the normalized current density $j_z = \mathrm{Re}(j_z(x) + j_z^{ext})$ for two values of $k/k_0$. It was assumed that external current density $j_z^{ext}$ did not depend on *x*. Actually this value determines the total current in the strip. These numerical results are obtained for a strip of width $w = 0.02/k_0$ and parameter $p \equiv dw/2p\,l^{\;2}$ =2. The last parameter is the ratio of the strip width to the transverse penetration length $l_{eff} = 2p\,l^2/d$. If p>>1, the dc current in the thin strip has the functional form of $[1-(2x-w)^2/w^2]^{-1/2}$ with maximum at the edges.

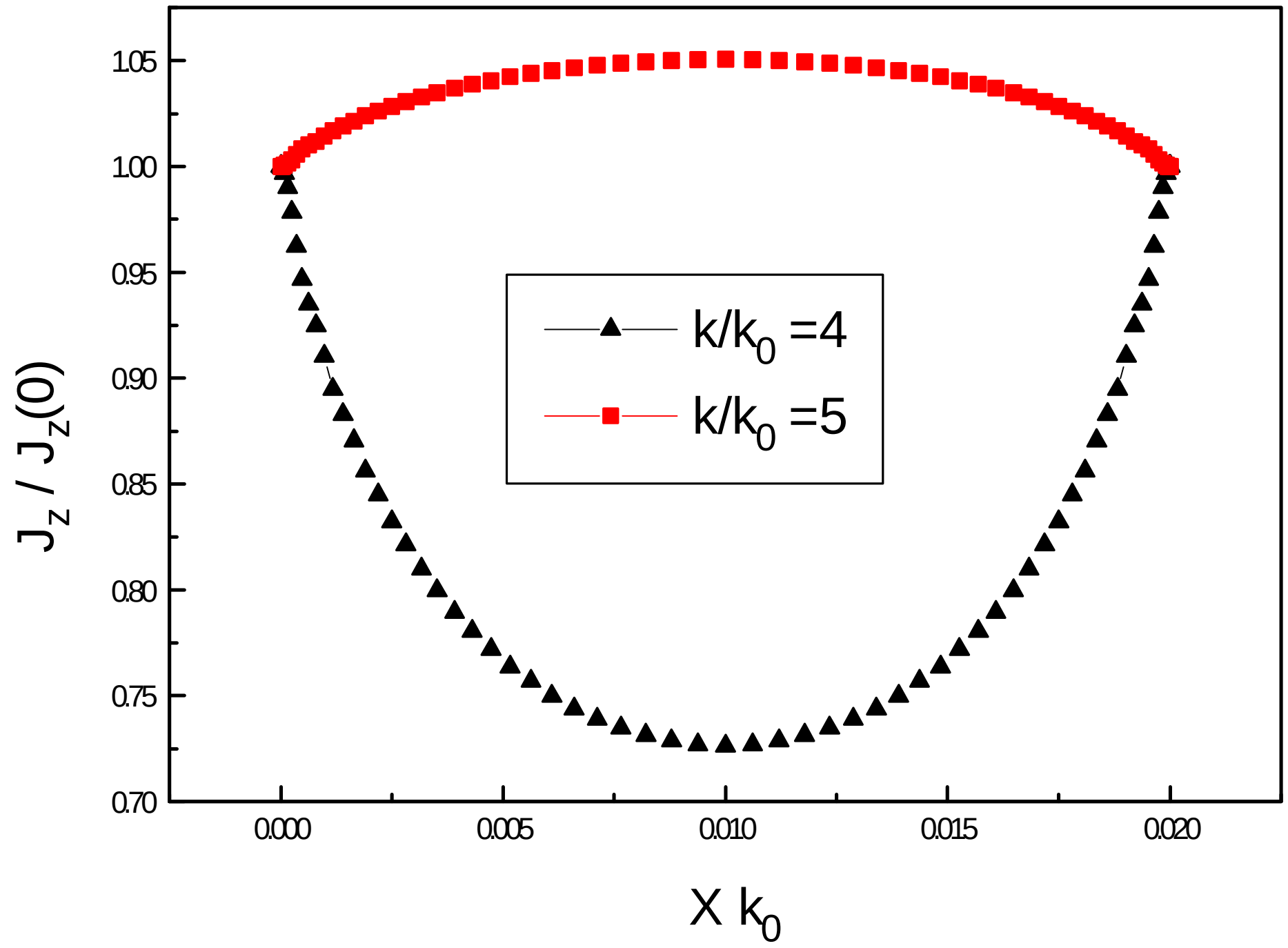

Fig. 3. Current pattern for two values of $k/k_0$. The stripline width is equal to $0.02/k_0$ and $p$=2.

We discussed up to now only the solutions of inhomogeneous Eq. (5) with $j_z^{ext} \neq 0$. The propagating microwave modes are determined by nontrivial solutions of homogeneous Eq.(5) with $j_z^{ext} = 0$. The dominant mode in this structure has $k/k_0 e^{1/2} > 1$ and, accordingly, current density has maximum over the center of the strip. For strip with $p$=2 the dominant mode has $k = 5.123k_0$. At Fig. 4 we showed the current patterns for the dominant mode for several values of parameter $p$. The increasing of $p$ leads to the decreasing the height of the maximum over the center of the strip. This is due to the decreasing of eigennumbers $k/k_0$ as $p$ increases. For example, for $p$=2 we have $k/k_0$ =5.123; and for $p$=20 $k/k_0$ =4.922. This decreasing of $k/k_0$ approaches the border where current pattern changes its character as it is shown at Fig. 3 and current distribution flattens.

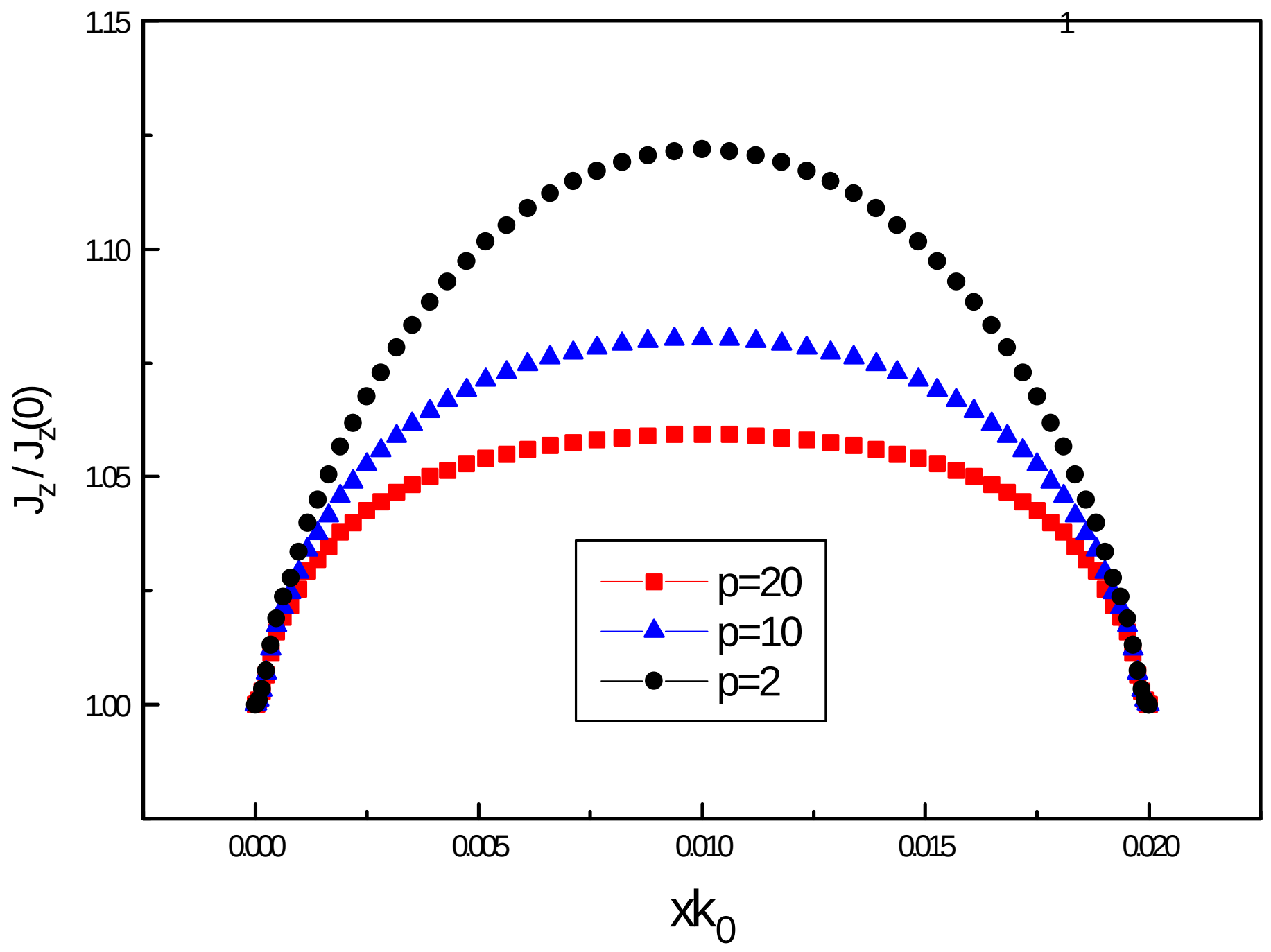

Fig. 4 Normalized current density for dominant propagating mode in striplines for several values of *p*.

The obtained results are different from published previous researches, which oversimplified the used model and neglected the dependence of the current distribution on *k*. Week's et al.[13] method exploited in [7,10,11] assumes that z-component of electric field is constant over the strip, but current density contradict the Ohm's law does not constant. Small parameter that validates this approach is not clear.

These numerical calculations take into account also $j_x$-component of the current and have been done in the following manner. The terms $j_z$ and $j_x$ are approximated by the expressions

$$j_z(x) = \sum_{0}^{N} a_n \cos(p\ \ nx/w)$$

$$j_x(x) = \sum_{1}^{N} b_n \sin(p\ \ nx/w) \qquad (8)$$

Because for a metal $rot\,\vec{H} = 4p\vec{j}\,/\,c$, then $div\,\vec{j} = 0$ and one could express $b_n$ through $a_n$ for n>0 .For term n=0 we have to use full equation $div\,\vec{j} + \partial r\,/\,\partial t\ = 0$, so the z-component of the total current($a_0$) in the strip does not equal zero(see discussion in the Appendix). A Galerkin testing procedure was applied in the space domain to yield the linear set of equations. Computation with harmonic number N=65 requires less than one minute of CPU time.

## Conclusion.

We found that current distribution in superconducting strip at microwave frequencies is different from distribution of steady current. There is maximum in the center of the strip. The increasing of the strip width leads to smoothing the current distribution.

## Appendix.

In this Appendix we discuss the application of equation $div\vec{j} = 0$ for calculation the wave propagation in a stripline. From exact equation

$$4p\ div\vec{j} + \partial div\vec{E}/\partial t\ = 0 \qquad \text{(A1)}$$

after integration over thickness of the strip we obtain

$$ds\,(\partial E_x/\partial x + \partial E_z/\partial z) + iw(E_y^+ - E_y^-)/4p = 0 \qquad \text{(A2)}$$

where $E_y^+(E_y^-)$ is the electric field on the top (bottom) surface of the strip and we took into account that $s >> w$ .From the z-component of equation $rot\vec{E} = ik_0\vec{H}$ we can obtain estimation $qE_y \approx \partial E_x/\partial y$ . If $q \approx n/w$>0 and $\partial E_x/\partial y \approx E_x/h$ then $E_y \approx E_x w/nh$, and proportional to ω term in (A1) can be neglected if $dsh/nw >> ww/nh$ . It means that we can use condition $div\vec{j} = 0$ for n >0 in (5). For n=0 we have to use (A1) if the total current in the strip does not equal zero.

## Literature.